\documentclass[
arial, showpacs,preprintnumbers,amsmath,amssymb]{revtex4}

\makeatletter

\newcommand{\Rmnum}[1]{\expandafter\@slowromancap\romannumeral #1@}
\makeatother

\usepackage{amssymb, amsmath, textcomp, dsfont, }
\usepackage{mathtools, verbatim, color}

\usepackage{graphicx}
\usepackage{dcolumn, verbatim}
\usepackage{bm}

\begin{document}

\title{A dynamic correspondence between Bose-Einstein condensates and Friedmann-Lema\^itre-Robertson-Walker and Bianchi I cosmology with a cosmological constant}

\author{Jennie D'Ambroise and Floyd L. Williams}
\affiliation{%
Department of Mathematics, University of Massachusetts, Amherst, Massachusetts 01003
}%

\date{ 21 April 2010}

\begin{abstract}
In some interesting work of James Lidsey, the dynamics of Friedmann-Lema\^itre-Robertson-Walker (FLRW) cosmology with positive curvature and a perfect fluid matter source is shown to be modeled in terms of a time-dependent, harmonically trapped Bose-Einstein condensate.  In the present work, we extend this dynamic correspondence to both FLRW and Bianchi I cosmologies in arbitrary dimension, especially when a cosmological constant is present.
\end{abstract}

\pacs{98.80.Jk, 03.75.Nt, 11.10.Kk,   04.20.Jb    }

\maketitle
\vspace{-.2in}
\begin{quote}
{\small \emph{
[Copyright (2010) American Institute of Physics. This article may be downloaded for personal use only. Any other use requires prior permission of the author and the American Institute of Physics.  The following article appeared in The Journal of Mathematical Physics {\bf 51} (2010), No. 6, 062501 and may be found at http://link.aip.org/link/JMAPAQ/v51/i6/p062501/s1. ]}
}
\end{quote}

\section{\label{sec:level1}INTRODUCTION\protect}

The general feature of this paper is a connection between a non-gravitational system and a gravitational system.  Such connections of course are of growing interest and importance.  More specifically, extending methods initiated in \cite{LidseyBEC}, we set up a correspondence between Bose-Einstein condensates governed by a time-dependent, harmonic trapping potential and both Friedmann-Lema\^itre-Robertson-Walker (FLRW) cosmology and Bianchi I cosmology, in an arbitrary dimension, and with a non-zero cosmological constant in both cases.  The correspondence is presented by way of two tables (Table I and Table II) that match cosmological parameters (scale factors, scalar pressure and energy density, Hubble parameter) with wavepacket parameters given in terms of the harmonic trapping frequency $\omega(t)$ and \emph{moments} $I_j(t), j=2, 3, 4$ (with $I_2(t)>0$) of the wavefunction of the Gross-Pitaevskii equation - under the assumption that the atomic interaction parameter is a constant in time.  Here $t$ is ``laboratory time" that one passes to from cosmic time $\tau$ in the Einstein field equations.  The moments satisfy the conservation law
\begin{equation}\lambda(t)\stackrel{def}{=}2I_2(t)I_4(t)-{I}_3^2(t)/4=\mbox{ a constant }\lambda.\end{equation}
Moreover, $I_2(t)$ gives rise to a solution $\Rmnum{10}(t)\stackrel{def.}{=}I_2^{1/2}(t)$ of the classical Ermakov-Milne-Pinney (EMP) equation \cite{Perez}
\begin{equation}\frac{d^2\Rmnum{10}}{dt^2}+\omega^2(t)\Rmnum{10}=\frac{\lambda}{\Rmnum{10}^3}.\end{equation}

On the other hand, one knows also that the Einstein equations for a FLRW universe, and even for the anisotropic Bianchi I and Bianchi V universes, admit a formulation in terms of a suitable EMP equation, of classical or of generalized type \cite{On3+1, JDFW, JDconfBI, JDthesis, HL, FWEMPBI, FWEFEFRW}.  This information coupled with equation (2) allows one to proceed in setting up the desired correspondence - at least in the FLRW and Bianchi I cases.

For certain cosmological models, for example a stiff fluid model, the frequency $\omega(t)$ can be determined, and thus the external potential is explicated.  A new issue arises however when a non-vanishing cosmological constant is present.  Our analysis shows that in this case one must employ elliptic functions to solve the appropriate moment equations that arise - a matter discussed in section V.

As indicated in \cite{LidseyBEC}, through the condensed matter - cosmology correspondence via EMP equations (as considered here) there is the increased possibility for insight into the hidden symmetries of these systems.  Further connections between non-gravitational and gravitational systems are explored in \cite{managedBEC}, for example.

\section{AN EMP FORMULATION OF FLRW AND BIANCHI I}

We begin by formulating the $d-$dimensional FLRW and Bianchi I equations with a cosmological constant $\Lambda_d$ (for $d\geq 3$) as a single EMP equation of classical type; we do this (for the record) as it has not been done before in arbitrary dimensions, even though the extension is straightforward.  The classical EMP equation is of the form 
\begin{equation}\ddot{Y}+A(t)Y=\frac{\mu}{Y^3}\end{equation}
for  $\ddot{Y}=\frac{d^2Y}{dt^2}$, and for some constant $\mu$; compare equation (2).  


We consider a scalar field $\phi$ and a potential $V$, and let $a(\tau)$ denote the scale factor, in which case the FLRW equations assume the following form, for the Hubble parameter $H\stackrel{def.}{=}\frac{a'}{a}:$
\begin{equation}H^2+\frac{k}{a^2}=\frac{2\Lambda_d}{(d-1)(d-2)}+\frac{2 K_d}{(d-1)(d-2)}\left[\frac{(\phi')^2}{2}+V\circ\phi+\frac{D}{a^n}\right],\end{equation}
\begin{equation}\phi'\phi''+(d-1)H(\phi')^2+(V'\circ\phi)\phi'=0,\end{equation}
with $k=0,-1,$ or $1$ the curvature parameter, $K_d\stackrel{def.}{=}8\pi G_d$ for $G_d=$ the gravitational constant and $\Lambda_d$ the cosmological constant.

Suppose $f(t)>0$ is a function with inverse function $T(\tau)$ (i.e. $f(T(\tau))=\tau, T(f(t))=t$) such that $T'(\tau)=a(\tau)$.  Define
\begin{equation}Y(t)\stackrel{def.}{=}a(f(t)), \phi_1(t)\stackrel{def.}{=}\phi(f(t)).\end{equation}
The method of \cite{HL} shows immediately that equations (4) and (5) lead to the EMP equation
\begin{equation}\ddot{Y}(t)+\frac{nK_d}{2(d-2)}\dot\phi_1(t)^2Y(t)=\frac{k}{Y(t)^{3}}.\end{equation}



For the anisotropic $d-$dimensional Bianchi I cosmological model with metric $ds^2=-d\tau^2+X_1(\tau)^2dx_1^2+\cdots+X_{d-1}(\tau)^2dx_{d-1}^2$,
 the field equations take the form
 \begin{equation}\displaystyle\sum_{l<k} H_lH_k=K_d\left[\frac{(\phi')^2}{2}+V\circ\phi\right]+\Lambda_d,\end{equation}
 \begin{equation}\displaystyle\sum_{l\neq i}\left(H_l'+H_l^2\right)+\displaystyle\sum_{\stackrel{l<k}{l,k\neq i}}H_lH_k=-K_d\left[\frac{(\phi')^2}{2}-V\circ\phi\right]+\Lambda_d\end{equation}
where $i,l,k\in\{1, \dots, d-1\}$ and $H_l\stackrel{def.}{=}\dot{X}_l/X_l$.  By making the substitution $X_l(\tau)=R(\tau)e^{\alpha_l(\tau)}$ for functions $R(\tau)>0$ and $\alpha_l(\tau)$ satisfying $\alpha_1(\tau)+\cdots +\alpha_{d-1}(\tau)=0$, one sees that the field equations (8) and (9) can be written as
\begin{equation}\frac{(d-1)(d-2)}{2}H_R^2-\frac{DK_d}{R^{2(d-1)}}=K_d\left[\frac{(\phi')^2}{2}+V\circ\phi\right]+\Lambda_d,\end{equation}
\begin{equation}(d-2)H_R'+\frac{(d-1)(d-2)}{2}H_R^2+\frac{DK_d}{R^{2(d-1)}}=-K_d\left[\frac{(\phi')^2}{2}-V\circ\phi\right]+\Lambda_d\end{equation}
since $R=(X_1X_2\cdots X_{d-1})^{1/(d-1)}$, where $ H_R(\tau)\stackrel{def.}{=}{R}'(\tau)/R(\tau)$, $D\stackrel{def.}{=}\frac{R^{2(d-1)}}{2(d-1)K_d}\sum_{l<k}(H_l-H_k)^2$.  $D$ is a constant quantity by a simple lemma which states that for any differentiable function $g(\tau)$, any positive differentiable function $R(\tau)$, and $M\in\mathds{R}$, the function $g(\tau)R(\tau)^M$ is constant if and only if $g'(\tau)+Mg(\tau)\frac{R'(\tau)}{R(\tau)}=0$.  By equating the left-hand sides of any two Einstein equations (9), indexed by $i\neq j$, we see that the latter equation holds for $g=H_i-H_j$ and $M=(d-1)$, so that $(H_i-H_j)R^{d-1}$ is constant for all $i,j\in\{1, \dots, d-1\}$.  


Similar to the above argument, by taking $f(t)>0$ to be the inverse of $T(\tau)$ such that $T'(\tau)=R(\tau)^{(d-1)}$, and by defining $Y(t)\stackrel{def.}{=}R(f(t))^{(d-1)}, \phi_1(t)\stackrel{def.}{=}\phi(f(t))$, we can show that the Bianchi I field equations (10), (11) lead to the classical EMP
\begin{equation}\ddot{Y}(t)+\frac{(d-1)K_d}{(d-2)}\dot\phi_1(t)^2Y(t)=\frac{-2(d-1)K_dD}{(d-2)Y(t)^3}.\end{equation}

\section{ELLIPTIC FUNCTIONS INTERLUDE}

In the case of a non-vanishing cosmological constant, that we shall give attention to, the following differential equation (which is of some independent interest)
\begin{equation}\frac{\dot{y}^2}{4}=\frac{2A}{y^2}-B+Cy\end{equation}
arises, where $A>0, C\neq 0$; see equation (35).  Its solutions involve elliptic functions.  Before setting up the dynamic correspondence between cosmological and condensate systems we present a brief interlude regarding equation (13), which in particular will allow, conveniently, for the introduction of some notation needed later.

We shall need the elliptic functions $EF(x,k), E(u,k)$ of the \emph{first} and \emph{second} kind, respectively, with modulus $k$ \cite{Akh, BF, Greenhill, Hancock, PS} given by
\begin{eqnarray}EF(x,k)&\stackrel{def.}{=}&\displaystyle\int_0^x\frac{dt}{\sqrt{(1-t^2)(1-k^2t^2)}},\\
E(u,k)&\stackrel{def.}{=}&\displaystyle\int_0^u dn^2(v,k)dv =\displaystyle\int_0^{sn(u,k)} \sqrt{\frac{1-k^2t^2}{1-t^2}}dt.\notag\end{eqnarray}
Sometimes, as is usual, the modulus $k$ is suppressed in the notation and one writes $E(u), sn(u)$, for example, for $E(u,k), sn(u,k)$.  Given that $C\neq 0$, we can define constants $m=m(C), n=n(B,C)$ and Weierstrass invariants $g_2=g_2(B,C), g_3=g_3(A,B, C), p=p(B,C), q=q(A, B, C)$ by 
\begin{eqnarray}
m&\stackrel{def.}{=}&\frac{C}{|C|}e^{\frac{1}{3}\log \frac{4}{|C|}} \qquad\mbox{( i.e. $m^3\stackrel{def.}{=}\frac{4}{C}$ )},\notag\\
n&\stackrel{def.}{=}&\frac{B}{3C}, \qquad g_2\stackrel{def.}{=}\frac{mB^2}{3C}=mnB,\\
g_3&\stackrel{def.}{=}&\frac{2B^3}{27 C^2}-2A, \qquad p\stackrel{def.}{=}-\frac{g_2}{4}=-\frac{mB^2}{12C},\notag\\
q&\stackrel{def.}{=}&-\frac{g_3}{4}=-\frac{B^3}{54C^2}+\frac{A}{2}.\notag\end{eqnarray}
The point of these definitions is that if $v(t)$ is the function given by $v(t)\stackrel{def.}{=}[y(t)-n]/m$, then equation (13) transforms to the differential equation
\begin{equation}\frac{\left(mv+\frac{B}{3C}\right)^2\dot{v}^2}{4(v^3+pv+q)} =\frac{4}{m^2},\end{equation}
as a direct computation reveals.  Moreover, if $w(x)$ is an inverse function of $y(t)$ then $z(x)\stackrel{def.}{=}w(mx+n)$ is an inverse function of $v(t)$ so that from equation (16) the equation 
\begin{equation}z'(x)^2=\frac{m^2}{16}\frac{\left(mx+\frac{B}{3C}\right)^2}{X(x)}, X(x)\stackrel{def.}{=}x^3+px+q\end{equation}
is directly derived.  This means that we can focus on solving
\begin{equation}z'(x)=\pm\left[\frac{mB}{12C}\frac{1}{\sqrt{X(x)}}+\frac{m^2}{4}\frac{x}{\sqrt{X(x)}}\right],\end{equation}
which involves the elliptic integrals $\displaystyle\int\frac{dx}{\sqrt{X(x)}}, \displaystyle\int\frac{xdx}{\sqrt{X(x)}}$, that in turn involve a consideration of the roots of the cubic equation $X(x)=0$ for their evaluation.  The case of interest here is when $X(x)=0$ has one real root $r_1$ and two complex roots $r_2, r_3$ - the other cases being simpler to deal with.  Necessarily $r_2$ and $r_3$ are complex conjugates: $r_3=\overline{r}_2$.  The condition that $X(x)=0$ indeed has a single real root $r_1$ and two complex conjugate roots $r_2,r_3$ is that its discriminant $\triangle=-4p^3-27q^2\stackrel{def.}{=}\frac{1}{16}\left[ g_2^3-27g_3^2\right]\stackrel{def.}{=}\frac{1}{16}\frac{4A}{C^2}\left[ 2B^3-27 AC^2\right]$ (since $m^3=\frac{4}{C}$) should be negative:  $2B^3-27AC^2<0$ (given that $A>0$), which we therefore assume.

Associated with the roots are useful parameters $\sigma, \rho, g, t_1, t_2:$
\begin{equation*}\sigma\stackrel{def.}{=}Im r_2,\quad \rho\stackrel{def.}{=}-r_1/2, \quad g\stackrel{def.}{=}\frac{1}{\sqrt[4]{9\rho^2+\sigma^2}},\end{equation*}
\begin{equation}t_1\stackrel{def.}{=} r_1+\sqrt{(\rho -r_1)^2+\sigma^2}=r_1+\sqrt{9\rho^2+\sigma^2}, \end{equation}
\begin{equation*}t_2\stackrel{def.}{=}r_1-\sqrt{(\rho-r_1)^2+\sigma^2}=r_1-\sqrt{9\rho^2+\sigma^2}.\end{equation*}
Here $\sigma\neq 0$, since otherwise $r_2$ would be a second real root.  The cubic $X(x)$ admits the factorization $X(x)=(x-r_1)[(x-\rho)^2+\sigma^2]$, which shows that (since $\sigma\neq 0$) $X(x)>0$ for $x>r_1$.  Thus $x>r_1$ will be a convenient assumption in our discussion of functions like $X(x)$ and $z'(x)$ in (18), for example, where $\frac{1}{\sqrt{X(x)}}$ appears.  The elliptic modulus $k$ that will be employed in (14) is given by
\begin{equation}k\stackrel{def.}{=}+ \sqrt{\frac{\sqrt{\frac{9r_1^2}{4}+\sigma^2}-\frac{3r_1}{2}}{2\sqrt{\frac{9r_1^2}{4}+\sigma^2}}}=+ \sqrt{\frac{\sqrt{9\rho^2+\sigma^2}+3\rho}{2\sqrt{9\rho^2+\sigma^2}}}.\end{equation}
We note that $0<k<1$.  For $\sigma\neq 0$,  $\sqrt{9\rho^2+\sigma^2}>\sqrt{9\rho^2}=3|\rho|\geq \pm 3\rho \Rightarrow 2\sqrt{9\rho^2+\sigma^2}=\sqrt{9\rho^2+\sigma^2}+\sqrt{9\rho^2+\sigma^2}>\sqrt{9\rho^2+\sigma^2}+3\rho>0$ so that $1>\frac{\sqrt{9\rho^2+\sigma^2} +3\rho}{2\sqrt{9\rho^2+\sigma^2}}>0$; i.e. $1>k^2>0\Rightarrow 1>k>0$, as desired.  

With the preceding notation and definitions in place, we can now construct a crucial function $u(x)=u(x,k)$ that facilitates the expression of solutions of equation (18).  For 
\begin{eqnarray}\theta(x)&\stackrel{def.}{=}&
1-\frac{(x-t_1)^2}{(x-t_2)^2}\notag\\
&=&\frac{[2x-(t_1+t_2)](t_1-t_2)}{(x-t_2)^2}, \quad r_1<x,
\end{eqnarray}
and for $EF(x,k)$ in (14) and $k$ in (20), we set
\begin{equation}u(x)=u(x,k)\stackrel{def.}{=}EF(\sqrt{\theta(x)},k).\end{equation}
Here $t_2<r_1<x\Rightarrow x-t_2\neq 0$ in (21).  Also $\theta(t_1)=1$, but for $r_1<x\neq t_1$ (i.e. $(x-t_1)^2>0$) $\theta(x)<1$.  Moreover, $2x>2r_1=t_1+t_2$ (by (19)) $\Rightarrow\theta(x)>0$ (since $t_1>t_2$).  That is, $r_1<x\neq t_1\Rightarrow 0<\theta(x)<1\Rightarrow 0<\sqrt{\theta(x)}<1$ in (22). Since $sn(y)$ is the inverse function of $EF(x)$, we can provide a second description of $u(x)$.  Namely, $\sqrt{\theta(x)}=sn(EF(\sqrt{\theta(x)}))\stackrel{\therefore}{=}sn(u(x))$, and therefore $cn(u(x))=\sqrt{1-sn^2(u(x))}=\sqrt{1-\theta(x)}\stackrel{def.}{=}\frac{|x-t_1|}{(x-t_2)}$ (since (again) $x>t_2$ for $x>r_1$) $\Rightarrow$
\begin{equation}u(x,k)=cn^{-1}\left(\frac{|x-t_1|}{x-t_2},k\right),\quad r_1<x.\end{equation}
In terms of the elliptic functions $EF(x,k), u(x,k)$ defined in (14) and (22) (or (23)), again with the elliptic modulus $k$ specified in definition (20), equation (18) is solved as follows, for an integration constant $z_0$:
\begin{equation}z(x)=\pm\left[\left(\frac{m^2}{2g}+\frac{m^2gt_2}{4}+\frac{mgB}{12C}\right)u(x)-\frac{m^2}{2g}EF(u(x))+\frac{m^2\sqrt{X(x)}}{2(x-t_2)}\right]+z_0\end{equation}
for $r_1<x<t_1$, and 
\begin{equation}z(x)=\pm\left[\left(-\frac{m^2}{2g}-\frac{m^2gt_2}{4}-\frac{mgB}{12C}\right)u(x)+\frac{m^2}{2g}EF(u(x))+\frac{m^2\sqrt{ X(x) }}{2(x-t_2)}\right]+z_0\end{equation}
for $t_1<x$.  $m, g, t_1, t_2$ are defined in (15), (19), and we assume that $A>0$ in (13), and $27AC^2>2B^3$ so that $r_1$ is the unique real root of the cubic equation $X(x)=0$, for $X(x)$ in (17) with $p,q$ there also defined in (15).  

Going back to the definition $z(x)\stackrel{def.}{=}w(mx+n)$ (for $n\stackrel{def.}{=}B/3C$ in (15)), where $w(x)$ is an inverse function of $y(t)$, we see that the initial differential equation (13) is solved implicitly by way of $w(x)$ given by
\begin{equation}w(x)=z\left(\frac{x-n}{m}\right),\end{equation}
for $z(x)$ given by (24), or by (25).  

\section{THE CORRESPONDENCES}

The Lidsey correspondence between BEC's and cosmology
%
originates by way of a comparison of equation (2) with equation (7), resulting in the following table.

\begin{table}[h]
\parbox{11cm}{
\caption{\label{tb: BEC-FLRW}BEC $\leftrightarrow$ FLRW correspondence 
}
}
\begin{tabular}{ccccc}\hline\hline\\
$I_2$ & & $\leftrightarrow$ && $a^2$\\
$I_3$ &&$\leftrightarrow$ && $2(aH)$\\
${I_3^2}/{4I_2}$ &&$ \leftrightarrow$ &&$ H^2$\\
$\left[ (d-1) (d-2) I_4-\Lambda_d \right]/K_d$ &&$ \leftrightarrow$ && $\rho_\phi$\\
$\left[(d-2)\omega^2I_2-(d-1)(d-2)I_4+\Lambda_d\right]/K_d$ && $\leftrightarrow $&&$ p_\phi$\\
\\
\hline\hline\end{tabular}
\end{table}


Similarly, the correspondence in the case of the Bianchi I cosmology originates by comparing equations (2) and (12) 
%
for which one obtains the following table.

\begin{table}[h]
\parbox{11cm}{
\caption{\label{tb: BEC-BI}BEC $\leftrightarrow$ Bianchi I correspondence 
}
}
\begin{tabular}{ccccc}\hline\hline\\
$I_2$ & &$ \leftrightarrow$ && $R^{2(d-1)}$\\
$I_3$ &&$\leftrightarrow$ &&$ 2(d-1)(R^{(d-1)}H_R)$\\
${I_3^2}/{4I_2}$ && $\leftrightarrow$ && $(d-1)^2H_R^2$\\
$\left[ \frac{(d-2)}{(d-1)} I_4-\Lambda_d \right]/K_d $&&$ \leftrightarrow $&& $\rho_\phi$\\
$\left[\frac{(d-2)}{(d-1)}\omega^2I_2-\frac{(d-2)}{(d-1)}I_4+\Lambda_d\right]/K_d$ &&$ \leftrightarrow$ && $p_\phi$\\
$\lambda $ &&$ \leftrightarrow $&&$ -2(d-1)K_dD/(d-2)$\\
\\
\hline\hline\end{tabular}
\end{table}


\section{SOME EXAMPLES}

The moments are known to satisfy the equations \cite{LidseyBEC, Perez}
\begin{eqnarray}\dot{I}_1(t)&=&0, \ \ \dot{I}_2(t)=I_3(t),\\
\dot{I}_3(t)&=&-2 \omega (t)^2 I_2(t)+4I_4(t), \ \  \dot{I}_4(t)=\frac{-\omega(t)^2}{2}I_3(t).\notag\end{eqnarray}
At this point we assume an equation of state $p_\phi=(\gamma-1)\rho_\phi$, with $\gamma>0$.  We indicate how to extend the discussion in section IV of \cite{LidseyBEC}.  The initial step is to equate the fifth BEC entry in Table I with $ \ (\gamma -1) \ $ times the forth entry there and solve for $\omega^2$.  The result is that 
\begin{equation}\omega^2=\gamma(d-1)\frac{I_4}{I_2}-\frac{\gamma\Lambda_d}{(d-2)I_2},\end{equation}
which by (27) gives (since $I_3=\dot{I}_2$)
\begin{equation}\dot{I}_4=\left[-\frac{\gamma(d-1)}{2}\frac{I_4}{I_2}+\frac{\gamma \Lambda_d}{2(d-2)I_2}\right]\dot{I}_2.\end{equation}
(29) being a first order, linear differential equation consequently has the solution 
\begin{equation}I_4=\frac{\alpha}{I_2^{\gamma(d-1)/2}}+\frac{\Lambda_d}{(d-1)(d-2)},\end{equation}
for an integration constant $\alpha$, which plugged into equation (28) yields the $\Lambda_d-$independent result
\begin{equation}\omega^2=\frac{\gamma(d-1)\alpha}{I_2^{[\gamma(d-1)+2]/2}}.\end{equation}
By equations (1) and (27)
\begin{equation}\lambda=2I_2I_4-I_3^2/4=2I_2I_4-\dot{I}_2^2/4,\end{equation}
which with the help of equation (30) can be re-written as 
\begin{equation}\frac{\dot{I}_2^2}{4}=\frac{2\alpha}{I_2^{[\gamma(d-1)-2]/2}}+\frac{2\Lambda_d I_2}{(d-1)(d-2)}-\lambda.\end{equation}

Note that the above assumption that $\gamma>0$ (so in particular, $\gamma\neq 0$) rules out the un-wanted conclusion $\omega=0$, by equation (31) (or (28)), and also the conclusion that $I_4(t)$ is a constant function, by equation (29).  Equations (31) and (33) govern the time-dependent trapping frequency and thus the external potential $V(r,t)=\omega(t)^2r^2/2$ also, although equation (33) is a bit complicated.  If $\Lambda_d=0$, for example, it has the implicit solution
\begin{equation}\frac{I_2^{q/2+1}{}_2F_1\left(\frac{1}{2}+\frac{1}{q}, \frac{1}{2}; \frac{3}{2}+\frac{1}{q}; \frac{\lambda 
I_2^q}{2\alpha}\right)}{\sqrt{2\alpha}(q+2)}=\pm t+t_0\end{equation}
for $q\stackrel{def.}{=}[\gamma(d-1)-2]/2$, and for an integration constant $t_0$.

Consider the choice $\gamma=6/(d-1)$, for example, which corresponds to a stiff perfect fluid.  Then equation (33) becomes
\begin{equation}\frac{\dot{I}_2^2}{4}=\frac{2\alpha}{I_2^2}-\lambda+\frac{2\Lambda_d I_2}{(d-1)(d-2)}.\end{equation}
This is equation (13) for $A\stackrel{def.}{=}\alpha, B\stackrel{def.}{=}\lambda$, and $C\stackrel{def.}{=}2\Lambda_d/(d-1)(d-2)$, with $\Lambda_d\neq 0$, where we note that $A=\alpha>0$ by (30), since $\gamma>0$.  Equation (35) can therefore be solved (implicitly) in terms of the elliptic functions $EF(x,k), E(u,k)$ in definition (14).  Namely, the inverse function of $I_2(t)$ is given by $w(x)=z\left(\frac{x-n}{m}\right)$ where $z(x)$ is given by equation (24) or (25), according to equation (26) and the notation of section III.  The hypothesis there is that $27 AC^2>2B^3$.  This means that $54\alpha\Lambda_d^2/(d-1)^2(d-2)^2>\lambda^3$, which as we have seen corresponds to the condition of a negative discriminant $\Delta$.  The cases $\Delta>0$ and $\Delta=0$ are treated in the Appendix.  By equation (31)
\begin{equation}\omega^2=\frac{6\alpha}{I_2^4}.\end{equation}

In case $\Lambda_d=0$, one can write equation (33) as $\frac{1}{2}I_2 dI_2/\sqrt{2\alpha-\lambda I_2^2}=\pm dt$, which integrated gives $-\frac{1}{2\lambda}\sqrt{2\alpha-\lambda I_2^2}=\pm t + t_0$, or 
\begin{equation}I_2^2(t)=\frac{2\alpha}{\lambda}-4\lambda\left[\pm t+t_0\right]^2,\end{equation}
for an integration constant $t_0$.  By equation (37), equation (36) is explicated.

Another choice of interest is $\gamma=\frac{4}{d-1}$, (for a universe dominated by matter when $d=4$) in which case equation (33) reads $\dot{I}_2^2/4=A/I_2-B+CI_2$, or 
\begin{equation}\displaystyle\int\frac{\sqrt{I_2}}{\sqrt{CI_2^2-BI_2+A}}dI_2=\pm 2t + t_0\end{equation}
where $A=2\alpha, B=\lambda, C=2\Lambda_d/(d-1)(d-2),$ and $t_0=$ an integration constant.  For a general value of $C$ the integral in (38) can be expressed explicitly in terms of the elliptic functions $EF(x,k), E(u,k)$ of the first and second kind in definition (14) by use of Maple, for example.  Thus again for $\Lambda_d\neq 0, I_2(t)$ can be determined implicitly, and moreover, by equation (31),
\begin{equation}\omega^2=\frac{4\alpha}{I_2^3}.\end{equation}

In the particular (easier) case when $\Lambda_d=0$ (i.e. $C=0$), for example, the integral in (38) is an elementary function.  Namely, equation (38) reduces to the equation 
\begin{equation}-\frac{\sqrt{A-BI_2}\sqrt{I_2}}{B}+\frac{A}{B^{3/2}}arctan\left(\sqrt{\frac{B I_2}{A-B I_2}}\right)=\pm 2t +t_0,\end{equation}
again for $A\stackrel{def.}{=}2\alpha >0, B\stackrel{def.}{=}\lambda$, say for $0< I_2(t)<\frac{A}{B}=\frac{2\alpha}{\lambda}, $ where already $I_2(t)>0$ and (as we have seen) $\alpha>0$.  Thus one needs that $\lambda>0$; in \cite{LidseyBEC} the choice $\lambda=1$ is made.

As a final example, regarding table I, we take $\gamma=\frac{3}{d-1}$.  Then equation (33) can be written as 
\begin{equation}\displaystyle\int\frac{I_2^{1/4}dI_2}{\sqrt{2\alpha+CI^{3/2}-\lambda I_2^{1/2}}}=\pm 2t+t_0,\end{equation}
again for $C\stackrel{def.}{=}2\Lambda_d/(d-1)(d-2)$.  However, the integral here is non-tractable unless $C=0$ (i.e. $\Lambda_d=0$), in which case its evaluation gives the implicit equation 
\begin{equation}\frac{6\alpha^2}{\lambda^{5/2}}arctan\left( \frac{\sqrt{\lambda} I_2^{1/4}}{\sqrt{2\alpha- \lambda I_2^{1/2}}}\right)-\sqrt{2\alpha - \lambda I_2^{1/2}}\left[\frac{3\alpha I_2^{1/4}}{\lambda^2}+\frac{I_2^{3/4}}{\lambda}\right]=\pm 2t+t_0\end{equation}
for $I_2(t)$.  Also for $\gamma=3/(d-1), \omega^2=3\alpha/I_2^{5/2}$ by equation (31).

To close things out we present a few examples regarding Table II for a Bianchi I cosmology, where we maintain the equation of state $p_\phi=(\gamma-1)\rho_\phi, \gamma>0$.  In place of equations (28) and (29), one quickly checks that the equations $\omega^2=\gamma I_4/I_2-\gamma(d-1)\Lambda_d/(d-2)I_2, \dot{I}_4+(\gamma/2)\dot{I}_2 I_4/I_2=(\gamma/2)(d-1)\Lambda_d \dot{I}_2/(d-2)I_2$ follow by Table II.  The latter equation has solution
\begin{equation}I_4=\frac{\alpha}{I_2^{\gamma/2}}+\left(\frac{d-1}{d-2}\right)\Lambda_d\end{equation}
by which the former equation can be written as 
\begin{equation}\omega^2=\frac{\gamma\alpha}{I_2^{\gamma/2+1}},\end{equation}
which again is $\Lambda_d$-independent, and where (again) $\alpha$ is an integration constant.  By equations (32) and (43) we deduce that 
\begin{equation}\frac{\dot{I}_2^2}{4}=\frac{2\alpha}{I_2^{\gamma/2-1}}+2\left(\frac{d-1}{d-2}\right)\Lambda_d I_2-\lambda\end{equation}
is the Bianchi I version of equation (33).  The equation 
\begin{equation}\displaystyle\int \frac{dI_2}{\sqrt{2\alpha I_2^{1-\gamma/2}+bI_2-\lambda}}=\pm 2 t+t_0\end{equation}
is a re-expression of equation (45) for $b\stackrel{def.}{=}2(d-1)\Lambda_d/(d-2)$.  

An obvious solution for $I_2$ is obtained by choosing $\gamma=2$, for example, in which case equation (46) reads (for $\Lambda_d\neq 0$) $(2/b)\sqrt{bI_2(t)-\lambda +2\alpha}=\pm 2t+t_0$, or 
\begin{equation}I_2(t)=\frac{\Lambda_d}{2}\left(\frac{d-1}{d-2}\right)(\pm 2t+t_0)^2+\frac{(\lambda-2\alpha)}{2\Lambda_d}\left(\frac{d-2}{d-1}\right).\end{equation}
Equation (44) then assumes the explicit form $\omega^2(t)=2\alpha/I_2^2(t)$.

As a second example, choose $\gamma=1$, $\Lambda_d=0$ (i.e. $b=0$).  Then equations (44) and (45) read 
\begin{equation}\frac{2\sqrt{2\alpha I_2^{1/2}(t)-\lambda \ }\left( \alpha I_2^{1/2}(t)+\lambda\right)}{3\alpha^2}=\pm 2t+t_0\notag\end{equation}
\begin{equation}\omega^2(t)=\alpha/I_2^{3/2}(t),\end{equation}
since $\displaystyle\int\frac{dx}{\sqrt{a\sqrt{x}-c}}=4\sqrt{a\sqrt{x}-c} \ (a\sqrt{x}+2c)/3a^2$.  More generally, the integral $\displaystyle\int\frac{dx}{\sqrt{ax^{1/(n+1)}-c}}$ for $n=0, 1, 2, 3 \cdots, $ which normally involves the hypergeometric function ${}_2F_1$, can be explicitly computed.  Thus for $\Lambda_d=0$, we obtain a family of examples by choosing $\gamma=\gamma_n\stackrel{def.}{=}2\left(1-\frac{1}{(n+1)}\right), n=0, 1, 2, 3 \dots$.  For the record, we are able to derive the general formula for the  corresponding integral in (46),
\begin{equation}\displaystyle\int\frac{dx}{\sqrt{a x^{1/(n+1)}-c}}=\frac{2(n+1)}{a^{n+1}}\sqrt{a x^{1/(n+1)}-c} \displaystyle\sum_{j=0}^n\frac{n!(ax^{1/(n+1)}-c)^{n-j}c^j}{j!(n-j)!(2n-2j+1)}.\end{equation}
The case $n=1$, was just treated.  If $n=2$, for example, $\gamma=\gamma_3=4/3$ and the corresponding integral in equation (46) is computed by the formula
\begin{equation}\displaystyle\int\frac{dx}{\sqrt{a x^{1/3}-c}}=\frac{2\sqrt{a x^{1/3}-c}}{5a^3}\left[ 3a^2x^{2/3}+4acx^{1/3}+8c^2\right].\end{equation}
If $n=3$, then $\gamma=\gamma_3=3/2$ and the integral in equation (46) is computed by 
\begin{equation}\displaystyle\int\frac{dx}{\sqrt{a x^{1/4}-c}}=\frac{8}{35a^4}\sqrt{a x^{1/4}-c} \left[
5a^3x^{3/4}+6a^2x^{1/2}c+8ax^{1/4}c^2+16c^3
\right].\end{equation}
In general for $\gamma=\gamma_n$, equation (44) shows that
\begin{equation}\omega^2=\frac{2n\alpha}{(n+1)I_2^{2-1/(n+1)}}.\end{equation}

\appendix

\section*{APPENDIX:  COMPUTATION OF SOME ELLIPTIC INTEGRALS}
\setcounter{section}{1}
\setcounter{equation}{0}

The problem of solving equation (13) has been reduced to that of solving equation (18), which in turn is a matter of computation of the elliptic integrals $I_j(x)\stackrel{def.}{=}\int\frac{x^j dx}{\sqrt{X(x)}}$.  For the reader's convenience we provide the result, which can be deduced from formulas in  \cite{BF} coupled with a few extra arguments.  We use freely the notation of section III.   

There are three cases: $\Delta<0, \Delta>0$, and $\Delta=0$.  First assume that $\Delta<0$.  That is,  $27AC^2>2B^3$ so that $X(x)=0$ has a single real root $r_1$.  Then, omitting integration constants, we have
\begin{eqnarray}
I_0(x)&=&\left[\begin{array}{cl}gu(x) & \mbox{for } r_1<x<t_1\\ -gu(x)& \mbox{for }t_1<x\end{array}\right],\notag\\
\end{eqnarray}
\begin{equation}I_1(x)=\left[
\begin{array}{cl}
\frac{2}{g}\left(u(x)-E(u(x))+\frac{g\sqrt{X(x)}}{x-t_2}\right)+t_2gu(x) & \mbox{for } r_1<x<t_1\\
-\frac{2}{g}\left(u(x)-E(u(x))-\frac{g\sqrt{X(x)}}{x-t_2}\right)-t_2gu(x) & \mbox{for }t_1<x\end{array}\right]\notag\end{equation}
for $u(x)=u(x,k)$ in definition (22).

If $\Delta>0$, i.e. $27AC^2<2B^2$, then $X(x)=0$ has three distinct \emph{real} roots $a,b,c$, say $a>b>c$.  In this case we now define $k,u(x)$ by
\begin{equation}k\stackrel{def.}{=}\sqrt{\frac{b-c}{a-c}}, \ u(x)=u(x,k)\stackrel{def.}{=}EF\left(\sqrt{\frac{x-a}{x-b}}, k\right).\end{equation}
Then $X(x)=(x-a)(x-b)(x-c)$ and for $x>a$
\begin{eqnarray}
I_0(x)&=&\frac{2u(x)}{\sqrt{a-c}},\notag\\
\end{eqnarray}
\begin{equation}\notag I_1(x)=2\sqrt{a-c}\left[dn \ u(x) tn \ u(x) - E(u(x))\right]+\frac{2au(x)}{\sqrt{a-c}},\end{equation}
where $tn \ x\stackrel{def.}{=}sn x / cn x$ as usual.  

If $\Delta=0$ (the final case), then $X(x)=0$ has at least two real roots: $X(x)=(x-a)^2(x-c)$ for real numbers $a,c$.  Thus the $I_j(x)$ are elementary functions computable by a calculus table of integrals, depending on whether $c=a$ or $c\neq a$.

\end{document}